\documentclass[twocolumn,secnumarabic,amssymb,nobibnotes,aps,superscriptaddress,pra]{revtex4-2} 
\usepackage[dvips]{graphicx}
\usepackage{bm}% bold math
\usepackage{latexsym} 
\usepackage{amsmath}
\usepackage{mathtools}
\usepackage{amssymb}
\usepackage{color}
\usepackage{ulem}
\usepackage {cancel}

\newcommand{\veps}{{\varepsilon}}
\newcommand{\eps}{{\epsilon}}

\newcommand{\bmA}{{\bm A}}

\newcommand{\bmr}{{\bm r}}
\newcommand{\bmp}{{\bm p}} 

\newcommand{\bmK}{{\bm K}}
\newcommand{\bmQ}{{\bm Q}}

\newcommand{\bmE}{{\bm E}}

\newcommand{\bms}{{\bm s}}

\newcommand{\bra}{\langle}
\newcommand{\ket}{\rangle}
\newcommand{\kB}{k_{\rm B}}

\makeatletter
\renewcommand*{\p@subsection}{}
\renewcommand*{\p@subsubsection}{}
\makeatother

\begin{document}

\title{Time-dependent Ginzburg-Landau theory of the vortex spin Hall effect}

\author{Hiroto Adachi}
\affiliation{Research Institute for Interdisciplinary Science, Okayama University, Okayama 700-8530, Japan}

\author{Yusuke Kato} 
\affiliation{Department of Basic Science, University of Tokyo, Meguro, Tokyo 153-8902, Japan}

\author{Jun-ichiro Ohe}
\affiliation{Department of Physics, Toho University, 2-2-1 Miyama, Funabashi 274-8510, Japan} 

\author{Masanori Ichioka}
\affiliation{Research Institute for Interdisciplinary Science, Okayama University, Okayama 700-8530, Japan} 
  
\date{\today}

\begin{abstract}
  We develop a time-dependent Ginzburg-Landau theory of the vortex spin Hall effect, i.e., a spin Hall effect that is driven by the motion of superconducting vortices. For the {\it direct} vortex spin Hall effect in which an input charge current drives the transverse spin current accompanying the vortex motion, we start from the well-known Schmid-Caroli-Maki solution for the time-dependent Ginzburg-Landau equation under the applied electric field, and find out the expression of the induced spin current. For the {\it inverse} vortex spin Hall effect in which an input spin current drives the longitudinal vortex motion and produces the transverse charge current, we microscopically construct the time-dependent Ginzburg-Landau equation under the applied spin accumulation gradient, and calculate the induced transverse charge current as well as the open circuit voltage. The time-dependent Ginzburg-Landau equation and its analytical solution developed here can be a basis for more quantitative numerical simulations of the vortex spin Hall effect. 
\end{abstract} 



\maketitle

%%%%%%%%%%%%%%%%%%%%%%%%%%%%%%%%%%%%
\section{Introduction \label{Sec:I}}
%%%%%%%%%%%%%%%%%%%%%%%%%%%%%%%%%%%%

Topological defect has been one of the important key concepts in condensed matter physics~\cite{Chaikin-textbook}. Recently, after the discovery of magnetic skyrmions in magnetic materials~\cite{Muehlbauer09,Yu10,Heinze11}, there has been a renewed interest in such real-space topological defects. Because of its topological robustness, the magnetic skyrmion is regarded as a useful information carrier~\cite{Nagaosa-review}. Besides magnetic systems, there is another well-known realization of real-space topological defects, that is a superconducting vortex~\cite{Fetter-review}. Indeed, as Bogdanov and Yablonskii pointed out a few decades ago~\cite{Bogdanov89}, there is a strong similarity between the magnetic skyrmions and the superconducting vortices, and one can observe their current-induced motion experimentally in both cases of magnetic skyrmions~\cite{Zeissler20} and superconducting vortices~\cite{Embon17}. Given this similarity as well as a great expectation for the use of the magnetic skyrmion as an information carrier~\cite{Tserkovnyak18}, it is natural to consider the possibility of transporting spin information by using the topological property of the superconducting vortices. 

\begin{figure}[t] 
  \begin{center}
    \includegraphics[width=8cm]{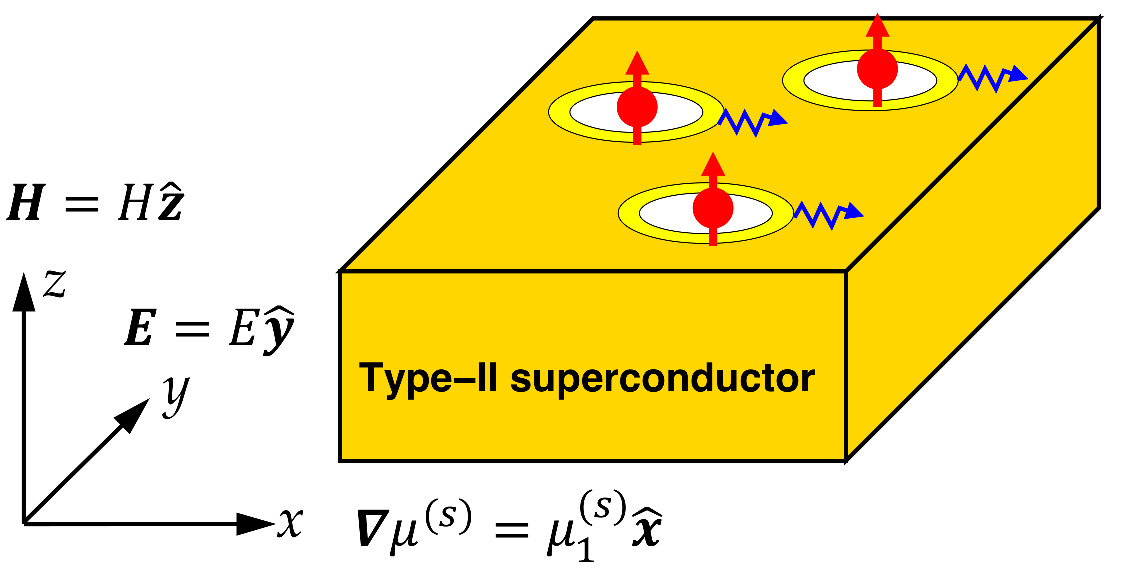}
  \end{center}
  \caption{ Schematic illustration of the vortex SHE. In the {\it direct} vortex SHE, an input electric field ${\bm E}= E{\bf \hat{y}}$ (or a charge current ${\bm j}^{\rm (c)}= \sigma_{\rm n} E {\bf \hat{y}}$) acts as the external force, which drives the vortex motion along the $x$ axis, carrying the spin current ${\bm J}^{\rm (s)} \parallel {\bf \hat{x}}$. In the {\it inverse} vortex SHE, an input spin accumulation gradient $\mu_1^{\rm (s)} {\bf \hat{x}}$ [or a spin current ${\bm j}^{\rm (s)}= (\sigma_{\rm n}/|e|) \mu_1^{\rm (s)}{\bf \hat{x}}$] acts as the external force, which drives the vortex motion along the $x$ axis, producing the charge current ${\bm J}^{\rm (c)} \parallel {\bf \hat{y}}$. Here, $\sigma_{\rm n}$ is the normal-state electrical conductivity and $e=-|e|$ is the electron charge. For more detail, see the main text. 
  }
  \label{fig:vSHE}
\end{figure}

Some years ago, along the line of the above argument, we theoretically investigated the vortex spin Hall effect (SHE), i.e., a novel SHE that is driven by the motion of superconducting vortices~\cite{Taira21}. The vortex SHE is an analog to the well-known vortex Ettingshausen/Nernst effect~\cite{Wang06,Pourret06}. In general, a spin-singlet Cooper pair does not host entropy, but a superconducting vortex does. Then, since the superconducting vortex moves approximately transverse to the charge current due to the Josephson equation~\cite{Josephson62,Anderson66}, the vortex motion is accompanied by a flow of the vortex core entropy, or a transverse heat current, producing the well-known vortex Ettingshausen/Nernst effect~\cite{Wang06,Pourret06}. In the case of the vortex SHE, the entropy held in the vortex core is replaced by the spin accumulation~\cite{Adachi05,Ichioka07}, giving rise to the vortex SHE (Fig.~\ref{fig:vSHE}). Note that a similar physical situation has been theoretically investigated in Refs.~\cite{Kim18} and \cite{Vargunin19}. In Ref.~\cite{Kim18} the angular momentum transport through a type-II superconductor in the form of {\it vorticity} is discussed based on the so-called spin-rotation coupling~\cite{Matsuo13,Takahashi16}. By contrast, in the present paper the transport in the form of {\it spin accumulation} trapped by vortices is discussed, such that the underlying physics is different. In Ref.~\cite{Vargunin19}, the physics same as the present paper is discussed, but the Keldysh-Usadel theory is used there. By contrast, here we formulate the vortex SHE in terms of the time-dependent Ginzburg-Landau (TDGL) theory, which has several advantages as emphasized after the next paragraph. 

In our previous publication~\cite{Taira21}, we computed the vortex spin Hall conductivity by using a diagrammatic calculation of the Kubo formula. Then, using the resultant spin Hall conductivity combined with the self-consistent Hartree approximation~\cite{Ullah91} of the superconducting fluctuations~\cite{Skocpol75,Larkin-textbook,Bray74,Thouless75,Ikeda89,Watanabe22}, we argued that this approach allows us to explain the characteristic temperature dependence of the voltage due to the {\it inverse} vortex SHE observed in a NbN/Y$_3$Fe$_5$O$_{12}$ bilayer system~\cite{Umeda18}. Note that a similar experiment using a NbN/Fe bilayer has recently been reported~\cite{Sharma23}, where the vortex SHE also plays an important role. Here, we would like to emphasize that the Kubo approach employed in Ref.~\cite{Taira21} is just one side of two equivalent approaches for the investigation of vortex transport phenomena. Namely, as was noted in the context of vortex Nernst effect~\cite{Ussishkin02}, the microscopic Kubo approach and the TDGL approach provide us with two equivalent descriptions of transport phenomena in the superconducting vortex state. Therefore, it is natural to expect that the TDGL theory of the vortex SHE should be developed. 

The TDGL equation of superconductivity is a dynamical equation that is designed to recover the static Ginzburg-Landau equation in thermal equilibrium~\cite{Schmid66,Caroli67}. The TDGL equation has several advantages over the microscopic Kubo approach. First, it has a high affinity to numerical simulation. Indeed, the TDGL equation has so far been used to simulate the phase transition dynamics~\cite{Frahm91,Kato91,Liu91} and transport properties~\cite{Al-Saidi03,Mukerjee04} of superconductors. Second, the TDGL equation allows us to explicitly write down the dynamical solution for the moving Abrikosov lattice~\cite{Maki69}, which provides us with an intuitive understanding of how the Abrikosov lattice moves under the action of an external electric field. Considering these advantages, the TDGL description of the vortex SHE is highly demanded. 

In this work, we develop a TDGL theory of the vortex SHE. For the investigation of the {\it direct} vortex SHE in which an input charge current drives the transverse spin current concomitant with the vortex motion, we use the established time-dependent Ginzburg-Landau equation under the applied electric voltage, and employ the Schmid-Caroli-Maki solution of the moving Abrikosov lattice~\cite{Schmid66,Caroli67}. With this known apparatus, we find out the expression of spin current carried by the vortex motion. For the description of the {\it inverse} vortex spin Hall effect in which an input spin current drives the longitudinal vortex motion and produces the transverse charge current, we first need to construct the TDGL equation under the applied spin accumulation gradient as there has been no such formulations. After formulating the TDGL equation under spin accumulation gradient, we accomplish the calculation of the induced transverse charge current as well as the voltage established under the open circuit condition.

The organization of this paper is as follows. In Sec.~\ref{Sec:II}, we define our model and present the procedure of microscopically constructing the TDGL equation in the presence of the spin accumulation gradient. In Sec.~\ref{Sec:VSHE} we theoretically describe the {\it direct} vortex SHE, while in Sec.~\ref{Sec:IVSHE} we investigate the {\it inverse} vortex SHE. Finally, in Sec.~\ref{Sec:VI}, we discuss and summarize our results. We use unit $\hbar=\kB=c=1$ throughout this paper. 

%%%%%%%%%%%%%%%%%%%%%%%%%%%%%%%%%%%%% 
%%%%%%%%%%%%%%%%%%%%%%%%%%%%%%%%%%%%

%%%%%%%%%%%%%%%%%%%%%%%%%%%%%%%%%%%%%%%%%%%%%%%
\section{TDGL equation under spin accumulation gradient \label{Sec:II}} 
%%%%%%%%%%%%%%%%%%%%%%%%%%%%%%%%%%%%%%%%%%%%%%%
In this section we first define our model Hamiltonian that describes a type-II superconductor under the spin accumulation gradient. Next, starting from this model Hamiltonian, we microscopically construct the TDGL equation under the influence of spin accumulation gradient, which is needed for the analysis of the {\it inverse} vortex SHE in Sec.~\ref{Sec:IVSHE}.

%%%%%%%%%%%%%%%%%%%%%%
\subsection{Model}
%%%%%%%%%%%%%%%%%%%%%%

We start from the following Hamiltonian for an $s$-wave superconductor in the dirty limit: 
%%%
\begin{eqnarray}
  {\cal H} &=& \sum_{\sigma} \int d^3 r \psi^\dag_\sigma (\bmr) \left[
    \frac{(-i {\bm \nabla} + |e| {\bm A} )^2}{2m} 
    - \mu_\sigma (\bmr) \right]
  \psi_\sigma (\bmr)
  \nonumber \\
  &+& {\cal H}_{\rm imp} + {\cal H}_{\rm BCS},  
  \label{eq:Htot_01}  
\end{eqnarray}
%%%
where ${\bm A}$ is the vector potential, $m$ and $|e|$ are the mass and absolute value of charge of an electron, $\psi_\sigma (\bmr)$ is the electron field operator for spin projection $\sigma=\pm$, and $\mu_\sigma$ is the spin-dependent chemical potential. The second term of Eq.~(\ref{eq:Htot_01}), 
%%%
\begin{equation}
  {\cal H}_{\rm imp} = \sum_{\sigma} \int d^3 r U(\bmr) \psi^\dag_\sigma (\bmr) \psi_\sigma (\bmr), 
\end{equation}
%%%
is the Hamiltonian for impurity potential $U(\bmr)= \sum_a U_{\rm imp} (\bmr- \bmr_a)$. After the impurity average denoted by $[ \cdots ]_{\rm av}$, the mean and variance of $U(\bmr)$ satisfy $[U(\bmr)]_{\rm av}=0$ and $[ U(\bmr)U(\bmr') ]_{\rm av}= (2 \pi N(0) \tau_{\rm imp})^{-1} \delta (\bmr- \bmr')$, where $N(0)$ and $\tau_{\rm imp}$ are the density of states per spin and electron lifetime, respectively. The third term of Eq.~(\ref{eq:Htot_01}), 
%%%
\begin{equation}
  {\cal H}_{\rm BCS} = - |g| \int d^3 r \Psi^\dag (\bmr)  \Psi (\bmr), 
\end{equation}
%%%
is the BCS Hamiltonian with the attractive interaction parameter $|g|$, where $\Psi(\bmr) = \psi_- (\bmr) \psi_+ (\bmr)$ is the pair field. Note that the Hamiltonian in Eq.~(\ref{eq:Htot_01}) does not contain any spin-orbit interactions. 

The spin-dependent chemical potential in Eq.~(\ref{eq:Htot_01}) can be separated into charge and spin components, 
%%%
\begin{equation}
  \mu_\sigma (\bmr) = \mu^{\rm (c)} (\bmr) + \frac{\sigma }{2} \mu^{\rm (s)} (\bmr),
  \label{eq:mus01}
\end{equation}
%%%
where $\mu^{\rm (c)} = (\mu_+ + \mu_-)/2$ is the electrochemical potential, and ${\mu}^{\rm (s)} = (\mu_+ - \mu_-)$ is the spin accumulation~\cite{Takahashi08}.
In this paper, we consider a situation where the electrochemical potential is spatially uniform as 
%%%
\begin{equation}
  \mu^{\rm (c)} (\bmr) = \mu_0^{\rm (c)},
  \label{eq:muc01}  
\end{equation}
%%%
but the spin accumulation varies on the scale of the spin diffusion length $\lambda^{\rm (s)}$, where $\lambda^{\rm (s)}$ is much longer than the inter-vortex spacing characterized by the magnetic length $\ell$, i.e., $\lambda^{\rm (s)} \gg \ell$. If we assume the spin accumulation $\mu^{\rm (s)}$ varies only along the $x$ axis, it is represented as 
%%%
\begin{equation}
  \mu^{\rm (s)}(\bmr) = {\mu}_0^{\rm (s)} + \mu_1^{\rm (s)} x,
  \label{eq:mus02}  
\end{equation}
%%%
where $\mu_0^{\rm (s)}$ is the spatially uniform part of $\mu^{\rm (s)}(\bmr)$, whereas $\mu_1^{\rm (s)}$ is its gradient. Note that, the spin accumulation gradient acts as an external force to drive the vortex motion and spin current along the $x$ axis. Experimentally, the spin accumulation in a superconductor is realized by injecting spins via the spin Seebeck effect~\cite{Umeda18,Sharma23} or the spin pumping~\cite{Inoue17,Yao18,Rogdakis19,Jeon20}. Substituting Eqs.~(\ref{eq:muc01}) and (\ref{eq:mus02}) into Eq.~(\ref{eq:mus01}), we obtain 
%%%
\begin{equation}
  \mu_\sigma (\bmr) = \overline{\mu}_\sigma + \sigma V^{\rm (s)}(\bmr), 
  \label{eq:mus03}  
\end{equation}
%%%
where $\overline{\mu}_\sigma= \mu_0^{\rm (c)}+ \sigma \mu_0^{\rm (s)}/2$ and $V^{\rm (s)} (\bmr)= \mu_1^{\rm (s)} x /2$. 

Now the spatially uniform part of spin accumulation $\overline{\mu}_\sigma$ can be absorbed into the single-particle Hamiltonian, and the spatially asymmetric part is regarded as an external force. This amounts to considering the following Hamiltonian: 
%%%
\begin{eqnarray}
  {\cal H} &=& {\cal H}_0 + {\cal H}_{\rm BCS} + {\cal H}_{\rm imp} + {\cal H}^{\rm (s)}_{\rm ext}, 
  \label{eq:H0_04}  
\end{eqnarray}
%%%
where
%%%
\begin{equation}
  {\cal H}_0 = \sum_{\sigma} \int d^3 r \psi^\dag_\sigma (\bmr) \left[
    \frac{(-i {\bm \nabla} + |e| {\bm A} )^2}{2m} 
    - \overline{\mu}_\sigma \right]
  \psi_\sigma (\bmr)
\end{equation}
%%%
coincides with the first term on the right-hand side of Eq.~(\ref{eq:Htot_01}) but $\mu_\sigma (\bmr)$ being replaced by $\overline{\mu}_\sigma$. The external Hamiltonian ${\cal H}^{\rm (s)}_{\rm ext}$ is given by 
%%% 
\begin{eqnarray}
      {\cal H}^{\rm (s)}_{\rm ext} &=& - \sum_{\sigma} \int d^3 r \, \sigma V^{\rm (s)} (\bmr) 
      \psi^\dag_\sigma (\bmr) \psi_\sigma (\bmr), 
  \label{eq:H_ext01}  
\end{eqnarray}
%%%
where $V^{\rm (s)}(\bmr)$ is defined below Eq.~(\ref{eq:mus03}).

%%%%%%%%%%%%%%%%%%%%%%%%%%%%%%%%%%%%% 
\begin{figure}[t] 
  \begin{center}
    \includegraphics[width=6.5cm]{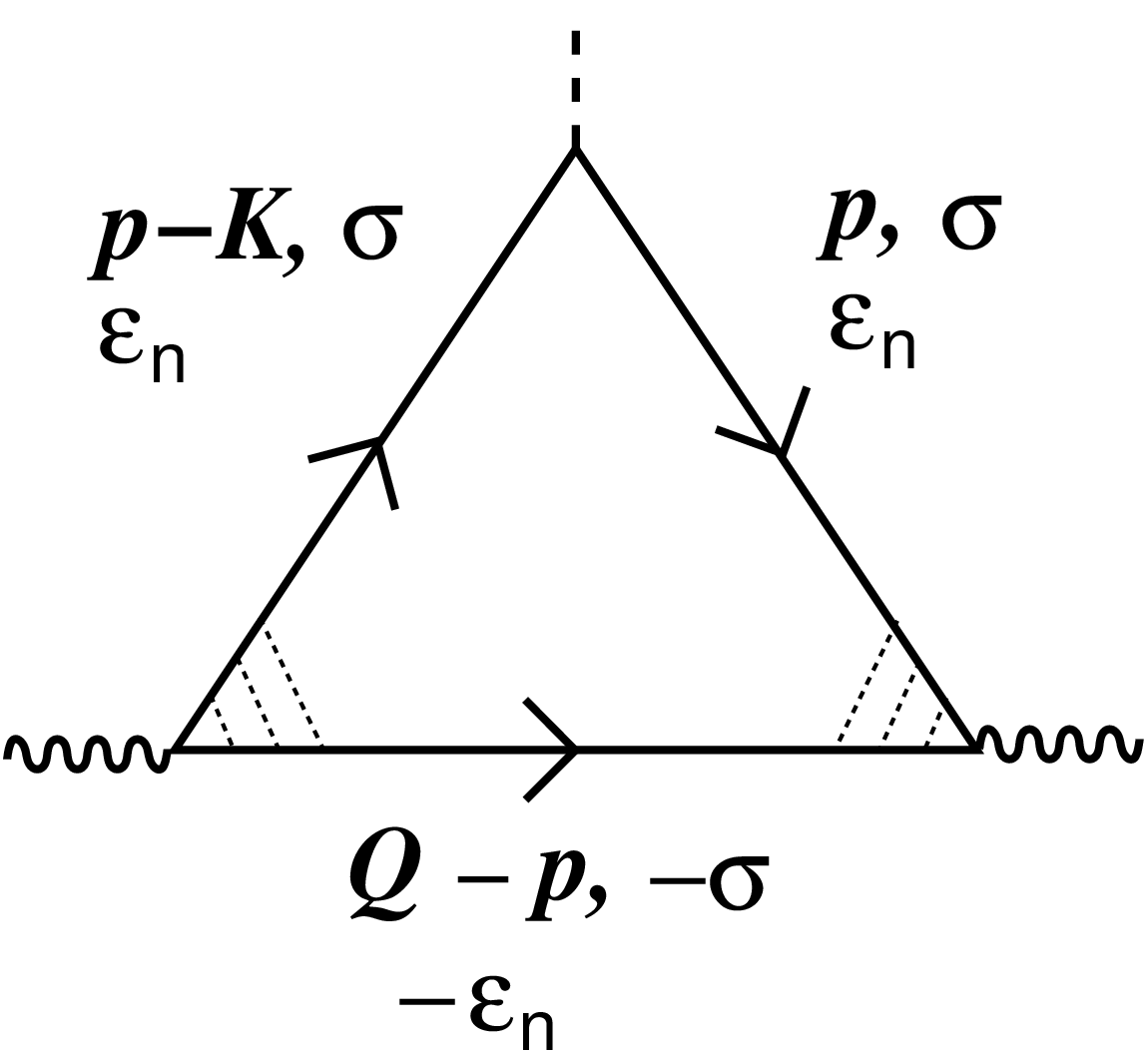}
  \end{center}
  \caption{Diagrammatic representation of Eq.~(\ref{eq:P02}), i.e., the first-order perturbation of particle-particle polarization in response to the spin accumulation gradient. A solid line with arrow is electronic Green's function, a dotted ladder is the Cooperon, a wavy line represents the pair field, and the dashed line represents the spin accumulation gradient [Eq.~(\ref{eq:spinaccum01})].
  }
  \label{fig:diagram02}
\end{figure} 
%%%%%%%%%%%%%%%%%%%%%%%%%%%%%%%%%%%%

%%%%%%%%%%%%%%%%%%%%%%%%%%%%%%%%%%%%%%%%%%%%%%%%%%%%%%%%%%%%%%%%%%%%%
\subsection{Effects of spin accumulation gradient on the TDGL equation} 
%%%%%%%%%%%%%%%%%%%%%%%%%%%%%%%%%%%%%%%%%%%%%%%%%%%%%%%%%%%%%%%%%%%%%
In this subsection, we microscopically construct the TDGL equation under the influence of the spin accumulation gradient. To this end, we first employ the formulation of microscopically constructing the TDGL equation in the presence of scalar potential gradient~\cite{Takayama70}, and then replace the scalar potential gradient with the spin accumulation gradient. 

The procedure of deriving the TDGL equation in the presence of the scalar potential gradient has been known in Ref.~\cite{Takayama70}, which is reviewed in Appendices~\ref{Sec:App01} and \ref{Sec:App02}. Our discussion here is based on the formulation therein. Then, the quantity necessary for investigating the coupling between the pair field and the scalar potential gradient is $\delta {\cal P}$ [Eq.~(\ref{eq:dltP01})], that is the first-order perturbation of particle-particle polarization in response to the scalar potential gradient represented by ${\cal H}^{\rm (c)}_{\rm ext}$ [Eq.~(\ref{eq:H_ext02})]. Now, in calculating $\delta {\cal P}$, we replace the scalar potential gradient ${\cal H}^{\rm (c)}_{\rm ext}$ with the spin accumulation gradient ${\cal H}^{\rm (s)}_{\rm ext}$ [Eq.~(\ref{eq:H_ext01})], and examine the first-order perturbation. In doing so, for the expression $V^{\rm (s)}(\bmr)$ appearing in ${\cal H}^{\rm (s)}_{\rm ext}$, we use 
%%%
\begin{equation} 
  V^{\rm (s)}(\bmr)= V^{\rm (s)}_\bmK e^{i \bmK \cdot \bmr}, 
  \label{eq:spinaccum01}
\end{equation}
%%%
where $V^{\rm (s)}_\bmK= (-i \mu_1^{\rm (s)}/2) \partial_{K_x}$ and the limit $\bmK \to 0$ is taken in the final step of the calculation. 

The resultant first-order perturbation of particle-particle polarization in response to ${\cal H}^{\rm (s)}_{\rm ext}$ is given by (see Fig.~\ref{fig:diagram02}), 
%%%
\begin{eqnarray}
  \delta {\cal P} &=& T \sum_\sigma \sum_{\veps_n} \int_\bmp
  G_{\bmp, \sigma}(\veps_n ) G_{\bmp-\bmK, \sigma}(\veps_n) G_{\bmQ-\bmp, -\sigma}(-\veps_n) 
  \nonumber \\
  &\times&   \left( -\sigma V_\bmK^{\rm (s)} e^{i \bmK \cdot \bmr }\right)
  \Big( {\cal C}_{\bmQ, \sigma}(\veps_n , -\veps_n) \Big)^2,  
  \label{eq:P02}
\end{eqnarray}
%%%
where we introduce the shorthand notation $\int_\bmp = \int d^3 p/(2 \pi)^3$. In the above equation,
%%%
\begin{eqnarray}
  G_{\bmp, \sigma}(\veps_n)
  &=& \left( i \widetilde{\veps}_n - \xi_\bmp+ \frac{\sigma}{2} \mu_0^{\rm (s)} \right)^{-1}
  \label{eq:Green01}
\end{eqnarray}
%%%
is the impurity-averaged Green's function, where $\widetilde{\veps}_n= \veps_n + {\rm sgn}(\veps_n)/(2 \tau_{\rm imp})$, $\xi_\bmp= p^2/2m- \mu_0^{\rm (c)}$, and $\veps_n= 2 \pi T (n+1/2)$ is a fermionic Matsubara frequency~\cite{Taira21}. Besides, 
%%%%%%%%%%%%%%%%%%%%%%%%
\begin{eqnarray}
  {\cal C}_{\bmQ,\sigma} (\veps_n+ \omega_m, -\veps_n)  \hspace{4.5cm} &&\nonumber \\
  = 
  \begin{cases}
    \frac{ \tau_{\rm imp}^{-1} }
         {d_\bmQ(2 \veps_n+ \omega_m) - i \sigma {\mu}_0^{\rm (s)} {\rm sgn}(\veps_n) }
         & \text{if $\veps_n(\veps_n+ \omega_m)>0 $,} \\
         1                 & \text{otherwise,} 
  \end{cases} &&
  \label{eq:Cooperon01}  
\end{eqnarray}
%%%%%%%%%%%%%%%%%%%%%%%%
is the Cooperon vertex, where $d_\bmQ(2 \veps_n+ \omega_m)= |2 \veps_n+ \omega_m|+ DQ^2$ with $D= v_F^2 \tau_{\rm imp}/3$ being the diffusion coefficient, and $\bmQ = -i{\bm \nabla} + 2|e|\bmA$. Note that, in contrast to the scalar potential gradient, the frequency dependence of the spin accumulation gradient in the present case can be safely neglected from the beginning. Consequently, diffuson does not appear in the above equation. 

After performing the momentum integral as well as the Matsubara frequency and spin summations, $\delta {\cal P}$ is calculated to be
%%%
\begin{eqnarray}
  \delta {\cal P} &=&
  \sum_{\sigma} \frac{i N(0) \sigma V^{\rm (s)}_\bmK e^{i \bmK \cdot \bmr} }{8 \pi T} \nonumber \\
  && \times 
  \left\{ \varPsi^{(1)} \left( \frac{1}{2}+ \frac{- i \sigma \mu_0^{\rm (s)}}{4 \pi T} \right)
  - (\sigma \leftrightarrow -\sigma) \right\},  
  \label{eq:P03} 
\end{eqnarray}
%%%
where $\varPsi^{(n)}(z)= d^n \varPsi(z)/d z^n$ with $\varPsi(z)$ being the digamma function, and a small correction $DQ^2$ is discarded. Then, expanding the result to the linear order in $\mu_0^{\rm (s)}$, we obtain 
%%%
\begin{eqnarray}
  \delta {\cal P} &=& N(0) \frac{ \mu_0^{\rm (s)} V^{\rm (s)}_\bmK e^{i \bmK \cdot \bmr}}{8 \pi^2 T^2}
  \varPsi^{(2)} \left(\frac{1}{2} \right), \nonumber \\
  &=&
  -N(0) \frac{ 7 \zeta (3) }{8 \pi^2 T^2} \mu_0^{\rm (s)} \mu_1^{\rm (s)} x,  
  \label{eq:dltP04s}
\end{eqnarray}
%%%
where in the last line we used $V^{\rm (s)}_\bmK e^{i \bmK \cdot \bmr} = \mu_1^{\rm (s)} x/2$, and $\varPsi^{(2)}(1/2) = - 14 \zeta (3) $ with $\zeta (z)$ being the zeta function. 

Substituting Eq.~(\ref{eq:dltP04s}) into Eq.~(\ref{eq:K2_02}), setting $\omega_m \to -i \omega= \partial_t$~\cite{Caroli67,Takayama70}, and using Eq.~(\ref{eq:TDGL00}), we finally obtain the TDGL equation under the spin accumulation gradient: 
%%%
\begin{equation}
  \left( \frac{\pi}{8T_c} \partial_t + \widetilde{\Lambda} x + \frac{T-T_c}{T_c}
  + \xi_0^2 \bmQ^2 + b |\Psi(\bmr,t)|^2 \right) \Psi(\bmr,t)= 0,
  \label{eq:TDGL02}
\end{equation}
%%%
where $\widetilde{\Lambda}= b \mu_0^{\rm (s)} \mu_1^{\rm (s)}$, and $b$ is defined below Eq.~(\ref{eq:TDGL01}). 

Equation~(\ref{eq:TDGL02}), i.e., the TDGL equation under the spin accumulation gradient, is formulated for the first time in the present paper. It will be used in Sec.~\ref{Sec:IVSHE} below for developing the TDGL theory of the {\it inverse} vortex SHE.

%%%%%%%%%%%%%%%%%%%%%%%%%%%%%%%%%%%%%%%%%%%%%%%
\section{TDGL theory of the direct Vortex spin Hall effect \label{Sec:VSHE}}  
%%%%%%%%%%%%%%%%%%%%%%%%%%%%%%%%%%%%%%%%%%%%%%%
In this section, we develop a TDGL theory of the {\it direct} vortex SHE in which an input electric field ${\bm E}= E{\bf \hat{y}}$ (or a charge current ${\bm j}^{\rm (c)}= \sigma_{\rm n} E {\bf \hat{y}}$) drives the transverse spin current ($\parallel {\bf \hat{x}}$) concomitant with the vortex motion, where $\sigma_{\rm n}$ is the normal-state electrical conductivity (see Fig.~\ref{fig:vSHE}). For this purpose, we employ the Schmid-Caroli-Maki solution of the moving Abrikosov lattice~\cite{Schmid66,Caroli67} for the TDGL equation under the applied electric voltage, and accomplish the calculation of the spin current carried by the vortex motion.

Note that the vortex spin current, i.e., the transverse spin current carried by the vortex motion that is driven by the longitudinal charge current, requires the spin polarization of the vortex core. Strictly speaking, such a situation can be realized in a flux flow state~\cite{Kim-review} with a small Zeeman splitting caused by an external magnetic field. But we believe a more efficient spin polarization is achieved by a spin injection into superconductors via the spin Seebeck effect~\cite{Umeda18,Sharma23} or the spin pumping~\cite{Inoue17,Yao18,Rogdakis19,Jeon20}, as these methods can inject spins locally into the vortex core in a site-selective way, causing minimal pair-breaking effect. Note also that, although the {\it direct} vortex SHE may in principle be observed by a Kerr rotation experiment~\cite{Kato04}, to the best of our knowledge there has been no report on the observation of the {\it direct} vortex SHE. Instead, the {\it inverse} vortex SHE which we discuss in the next section has been observed in Refs.~\cite{Umeda18,Sharma23}.

We begin with the following {\it linearlized} time-dependent Ginzburg-Landau equation, whose microscopic derivation is reviewed in Appendices \ref{Sec:App01} and \ref{Sec:App02} [see Eq.~(\ref{eq:TDGL01c})]: 
%%%
\begin{equation}
  \left( \partial_t  - 2 i |e| V^{\rm (c)}(\bmr) + D \bmQ^2 + \eps \right)  \Psi(\bmr,t) = 0,
  \label{eq:TDGL03}
\end{equation}
%%%
where $V^{\rm (c)}$ is the scalar potential, $D$ is the diffusion coefficient, $\bmQ$ is the gauge-invariant gradient defined below Eq.~(\ref{eq:gauge-grad01}), and $\eps= 8(T-T_{c})/\pi$ with $T_{c}$ being the superconducting transition temperature in the mean-field approximation under zero magnetic field. We choose the gauge ${\bm A}= H x {\bf \hat{y}}$ and $V^{\rm (c)}= - E y$, where ${\bm H}= H {\bf \hat{z}}$ is the external magnetic field and ${\bm E}= E {\bf \hat{y}}$ is the applied electric field. Note that the scalar potential appears in a gauge-invariant manner, i.e., $\partial_t  - 2 i |e| V^{\rm (c)}$. Note also that, in the present case in the absence of ${\mu}_1^{\rm (s)}$, there is no correction to $\epsilon$ within the linear order with respect to ${\mu}_0^{\rm (s)}$. 

Following Schmid~\cite{Schmid66}, Caroli and Maki~\cite{Caroli67}, we construct a flux-flow solution of Eq.~(\ref{eq:TDGL03}), 
%%%
\begin{eqnarray}
  \Psi(\bmr,t) &=& \sqrt{\frac{k_0 \ell  \langle |\Psi|^2 \rangle_{\rm s} }{\sqrt{\pi}}}
  \sum_{p=-\infty}^{\infty} C_p f_p(\bmr,t), \label{eq:Schmid01} \\
  f_p (\bmr,t) &=&
  \exp\left\{-\frac{(x+ k_0 p \ell^2 - ut)^2}{2l^2} \right\} \nonumber \\
  && \times   \exp\left\{i \left( k_0 p - \frac{ut}{\ell^2}+ \frac{v}{D} \right)y \right\}, 
  \label{eq:Schmid02} 
\end{eqnarray}
%%%
where $u$ and $v$ are parameters to be determined below, $\ell= 1/\sqrt{2|e|H_{\rm c2}}$ is the magnetic length at the upper critical field $H_{\rm c2}$, and $\langle |\Psi|^2 \rangle_{\rm s}$ means the spatial average of $|\Psi|^2$. Here, we use $k_0= \sqrt{\sqrt{3}\pi}/\ell$ and $C_p= e^{i \pi p^2/2}$ in order to reproduce the triangular Abrikosov lattice in the absence of $\bmE$.
  
Now, using $Q_x= -i \partial_x$ and $Q_y= -i \partial_y + x/\ell^2$ in the present gauge, we obtain 
%%%
\begin{subequations} \label{all}
  \begin{align}
    \partial_t f_p &= \left[ \frac{u (x+k_0 p \ell^2)}{\ell^2}
      - i \frac{uy}{\ell^2} \right] f_p, 
    \label{eq:DtDf01} \\
    Q_x^2 f_p &= \left[  \frac{1}{\ell^2}
      - \frac{(x+k_0 p \ell^2 -ut)^2}{\ell^4} \right] f_p,
    \label{eq:DQxDf01}\\
    Q_y^2 f_p &= \left[ \frac{(x+k_0 p \ell^2 -ut)^2}{\ell^4}
      + \frac{2v(x+ k_0 p \ell^2)}{D\ell^2}  \right] f_p , \; 
    \label{eq:DQyDf01}
  \end{align}
\end{subequations}
%%%
where only terms up to the linear order with respect to $u$ and $v$ are collected, except for terms appearing in the $(x+k_0 p \ell^2 -ut)^2$ combination. Then, substituting Eqs.~(\ref{eq:DtDf01})-(\ref{eq:DQyDf01}) to Eqs.~(\ref{eq:TDGL03}) and (\ref{eq:Schmid01}), we have 
%%%
\begin{eqnarray}
  && \Bigg\{ \left( \frac{u}{\ell^2}+ \frac{2v}{\ell^2}\right) (x+ k_0 p \ell^2) \hspace{2cm} \nonumber \\
  && \qquad 
  + i \left( \frac{E}{H_{\rm c2} \ell^2}- \frac{u}{\ell^2} \right) y
  + \eps+ \frac{D}{\ell^2} \Bigg\}
  f_p = 0. 
\end{eqnarray}
%%%
From the above equation we find that, $\Psi$ given in Eq.~(\ref{eq:Schmid01}) satisfies the TDGL equation (\ref{eq:TDGL03}) in the immediate vicinity of the upper critical field $H_{\rm c2}$ determined by $\eps+ D/\ell^2=0$, when 
%%%
\begin{eqnarray}
  u &=& \frac{E}{H_{\rm c2}}, \label{eq:u01}\\
  v &=& - \frac{u}{2}. \label{eq:v01}
\end{eqnarray}
%%% 

Since we have obtained the solution to the TDGL equation (\ref{eq:TDGL03}), we now check the expression of charge current ${\bm J}^{\rm (c)}$ carried by the vortex motion. Calculation of this quantity is basically the same as that given in Ref.~\cite{Maki69}, and here we only give a brief summary. The charge current driven by the vortex motion is given by~\cite{Ussishkin02,Maki69} 
%%%
\begin{equation}
  {\bm J}^{\rm (c)} = - K^{\rm (c)} \Psi^* \bmQ \Psi + \text{\rm c.c.}, 
\end{equation}
%%%
where $K^{\rm (c)}= 2|e|N(0) \xi_0^2$, and $\xi_0^2= \pi D/8 T_{c}$ was defined below Eq.~(\ref{eq:TDGL01}). By using the expression of $\Psi$ in Eq.~(\ref{eq:Schmid01}), we obtain 
%%%
\begin{eqnarray}
  J_x^{\rm (c)}
  &=&
  - K^{\rm (c)} \nabla_y |\Psi|^2 
\end{eqnarray}
%%%
and
%%%
\begin{eqnarray}
  J_y^{\rm (c)} 
  &=&
  K^{\rm (c)} \nabla_x |\Psi|^2 + K^{\rm (c)} \frac{u}{D} |\Psi|^2, 
\end{eqnarray}
%%%
which can be summarized into 
%%%
\begin{equation}
  {\bm J}^{\rm (c)} = {\bm \nabla} \times \left( - K^{\rm (c)}
  |\Psi|^2 {\bf \hat{z}} \right)
  + \frac{ K^{\rm (c)} u}{D} |\Psi|^2 {\bf \hat{y}}. 
  \label{eq:jc01}
\end{equation}
Then, after the spatial average, we obtain 
%%%
\begin{equation}
  \bra {\bm J}^{\rm (c)} \ket_{\rm s} =
  \frac{ K^{\rm (c)} u}{D} \bra |\Psi|^2 \ket_{\rm s} {\bf \hat{y}},
  \label{eq:jc02}
\end{equation}
where the first term on the right-hand side of Eq.~(\ref{eq:jc01}) vanishes upon the spatial average.

Experimentally, we observe the following total charge current: 
%%%
\begin{eqnarray}
  \bra {\bf J}_{\rm tot}^{\rm (c)} \ket_{\rm s}
  &=& \sigma_{\rm n} \left( E+  \frac{ K^{\rm (c)} u}{\sigma_{\rm n} D} \bra |\Psi|^2 \ket_{\rm s} \right) {\bf \hat{y}} \nonumber \\
  &=& \sigma_0 E {\bf \hat{y}},
  \label{eq:Jc_tot01}
\end{eqnarray}
%%%
where $\sigma_{\rm n}$ is the normal state electrical conductivity,
%%%
\begin{equation}
  \sigma_0 = \sigma_{\rm n} (1+ \eta)
\end{equation}
%%%
is the total electrical conductivity, and the dimensionless parameter $\eta$ is defined by
%%%
\begin{equation}
  \eta = \frac{1}{D \sigma_{\rm n}} \frac{K^{\rm (c)}\bra |\Psi|^2 \ket_{\rm s} }{H_{\rm c2}}.
  \label{eq:eta01}
\end{equation}
%%%   
Note that the electrical conductivity $\sigma_0$ is increased by a growth of the pair-field correlation $\bra |\Psi|^2 \ket_{\rm s}$ below $H_{\rm c2}$, whereas the electrical resistivity $\rho_0= 1/\sigma_0$ is decreased. 

Now, we are in a position to present the calculation of the spin current ${\bm J}^{\rm (s)}$ carried by the vortex motion. This quantity can be calculated from ~\cite{Taira21,Ussishkin02} 
%%%
\begin{eqnarray}
  {\bm J}^{\rm (s)} = - K^{\rm (s)}
  \Big[ \big( -i \partial_t + 2|e| V^{\rm (c)} \big) \Psi^* \Big]
  \bmQ \Psi + \text{\rm c.c.},
  \label{eq:js01}
\end{eqnarray}
%%%
where $K^{\rm (s)}= 2|e| \mu_0^{\rm (s)} N(0) \xi_0^2/(8 T_c^2)$, the time derivative in the above equation is understood to operate only on $\Psi^*$, and we use the relation between ${\bm J}^{\rm (s)}$ and the heat current ${\bm J}^{\rm (h)}$, i.e., ${\bm J}^{\rm (s)} = -2|e| \mu^{\rm (s)}_0 {\bm J}^{\rm (h)}/(8 T_c^2)$~\cite{Taira21}. 
Then, by using the expression of $\Psi$ in Eq.~(\ref{eq:Schmid01}), we obtain
%%%
\begin{eqnarray}
  J_x^{\rm (s)} &=& - 2 u K^{\rm (s)}
  \sum_{p,p'} C^*_p f^*_p C_{p'} f_{p'} \nonumber \\
  && \quad \times \left[
    \frac{(x+ k_0 p \ell^2)(x+ k_0 p' \ell^2)}{\ell^4} \right], 
\end{eqnarray}
%%%
where higher order terms with respect to $u$ have been discarded, and we used
%%%
\begin{equation}
  \big( -i \partial_t + 2|e| V^{\rm (c)} \big) f_p^*
  =
  -i u \frac{(x+k_0 p \ell^2)}{\ell^2} f_p^*
\end{equation}
%%%
and 
%%%
\begin{equation}
  Q_x f_{p'} 
  =
  i\frac{(x+k_0 p' \ell^2 - ut)}{\ell^2} f_{p'}.
\end{equation}
%%%
In a similar manner, we get 
%%%
\begin{eqnarray}  
  J_y^{\rm (s)} &=& \frac{-iu}{\ell^4} \sum_{p,p'} C^*_p C_{p'} (x+k_0p \ell^2)(x+k_0p' \ell^2) f^*_p f_{p'}
  + \text{c.c.} \nonumber \\
  &=& 0, 
\end{eqnarray}
%%%
where we used
%%%
\begin{equation}
  Q_y f_{p'} 
  =
  \frac{1}{\ell^2} \left( x+k_0 p' \ell^2 - ut - \frac{u \ell^2}{2D} \right) f_{p'}.
\end{equation}
%%%
Then, after the spatial average, we have 
%%%
\begin{equation}
  \bra {\bm J}^{\rm (s)} \ket_{\rm s} = - u \frac{K^{\rm (s)}}{\ell^2} \bra |\Psi|^2 \ket_{\rm s} {\bf \hat{x}},
  \label{eq:js02}
\end{equation}
%%%
where we used 
%%%
\begin{equation}
  \left\langle \sum_{p,p'} (x+k_0p \ell^2)(x+k_0p' \ell^2) f^*_p f_{p'} \right\rangle_{\rm s} 
  =
  \frac{\ell^2}{2} \langle |\Psi|^2 \rangle_{\rm s}. 
\end{equation}
%%%
The minus sign of $\bra {\bm J}^{\rm (s)} \ket_{\rm s}$ relative to $u$ comes from the definition of the spin current [Eq.~(\ref{eq:js01})]. Finally, we define the vortex spin Hall angle as $\Theta_{\rm VSHE} = |\bra {\bm J}^{\rm (s)} \ket_{\rm s}|/|\bra {\bf J}_{\rm tot}^{\rm (c)} \ket_{\rm s}|$. Then, we obtain 
%%%
\begin{equation}
  \Theta_{\rm VSHE} = \frac{\mu_0^{\rm (s)}}{\pi T_c} \left( \frac{\eta}{1+ \eta} \right), 
\end{equation}
%%%
by taking the ratio of Eqs.~(\ref{eq:Jc_tot01}) and (\ref{eq:js02}).

To summarize this section, we have developed a TDGL theory of the {\it direct} vortex SHE. In particular, we have obtained the explicit expression of the induced spin current within the TDGL theory, Eq.~(\ref{eq:js02}), which has not been known so far. This phenomenon, in which an input electric field ${\bm E}= E{\bf \hat{y}}$ (or a charge current ${\bm j}^{\rm (c)}= \sigma_{\rm n} E {\bf \hat{y}}$) drives the transverse spin current ($\parallel {\bf \hat{x}}$) concomitant with the vortex motion, can be viewed as a new member of the {\it direct} SHE, because the phenomenology matches the definition of the {\it direct} SHE~\cite{Takahashi08,Sinova15}.

%%%%%%%%%%%%%%%%%%%%%%%%%%%%%%%%%%%%%%%%%%%%%%%
\section{TDGL theory of the inverse vortex spin Hall effect \label{Sec:IVSHE}}  
%%%%%%%%%%%%%%%%%%%%%%%%%%%%%%%%%%%%%%%%%%%%%%%

In this section, we develop a TDGL theory of the {\it inverse} vortex SHE in which an input spin accumulation gradient $\mu_1^{\rm (s)} {\bf \hat{x}}$ [or a spin current ${\bm j}^{\rm (s)}= (\sigma_{\rm n}/|e|) \mu_1^{\rm (s)}{\bf \hat{x}}$] acts as the external force, which drives the vortex motion along the $x$ axis, producing the charge current ( $\parallel {\bf \hat{y}}$, see Fig.~\ref{fig:vSHE}). The TDGL equation with a spin accumulation gradient as the driving force was formulated in Sec.~\ref{Sec:II}. Below, starting from this TDGL equation, we calculate the induced transverse charge current as well as the voltage established under the open circuit condition. 

We begin with the {\it linearlized} TDGL equation under the spin accumulation gradient derived in Sec.~\ref{Sec:II} [Eq.~(\ref{eq:TDGL02})]: 
%%%
\begin{equation}
  \Big( \partial_t  - 2 i |e| V_{\rm iVSHE} + \Lambda x+ D \bmQ^2 + \eps \Big)  \Psi(\bmr,t) = 0, 
  \label{eq:TDGL04}
\end{equation}
%%%
where $\Lambda$ is given by 
%%%
\begin{equation}
  \Lambda = \frac{7 \zeta(3)}{\pi^3 T_c} \mu^{\rm (s)}_0 \mu^{\rm (s)}_1 .
  \label{eq:Lambda01}
\end{equation}
%%%
Note that the direction of the spin accumulation gradient in the above TDGL equation is arbitrary. But in order to explicitly write down the solution to the TDGL equation {\it in the specific gauge of} ${\bm A}= H x {\bf \hat{y}}$, choosing $x$ axis as the direction of spin accumulation gradient is convenient, which we adopt below. Note also that, although no electric voltage is applied to the system, because the vortex motion inevitably induces a transverse voltage due to the Josephson equation~\cite{Josephson62,Anderson66}, we find that a voltage caused by the {\it inverse} vortex SHE, $V_{\rm iVSHE}= -E_{\rm iVSHE} \, y$, is necessary for a consistent description of the {\it inverse} vortex SHE. 

Since the direction of the vortex motion is the same as in the previous section, we try to find a solution to Eq.~(\ref{eq:TDGL04}) in a form similar to Eq.~(\ref{eq:Schmid01}). Then, we find that, upon replacing $f_p(\bmr,t)$ with $g_p (\bmr,t) = f_p (\bmr,t) \exp(\Lambda k_0 p \ell^2 t)$, we can solve Eq.~(\ref{eq:TDGL04}) after choosing appropriate values of $u$ and $v$. Because of the additional factor $\exp(\Lambda k_0 p \ell^2 t)$ in $g_p$, however, the resultant $|\Psi(\bmr,t)|^2$ does not keep the periodic structure of the Abrikosov lattice. Therefore, we conclude that the spin accumulation gradient tends to break the vortex lattice structure, and it is impossible to construct the moving Abrikosov lattice solution to the {\it inverse} vortex SHE. 

The above consideration implies that we cannot construct a solution to Eq.~(\ref{eq:TDGL04}) that keeps the vortex lattice structure. Therefore we relax the constraint and try to find a solution to Eq.~(\ref{eq:TDGL04}) in the following form: 
%%%
\begin{eqnarray}
  \Psi(\bmr,t) &=& \sqrt{\frac{2 \sqrt{\pi \ell^2}}{L_y}}
  \sum_{k} \phi_k(t) u_k (\bmr,t) , \label{eq:iVSHE01}\\
  u_k (\bmr,t) &=& 
  \exp\left\{-\frac{(x+ k \ell^2 - ut)^2}{2l^2} \right\} \nonumber \\
  && \times   \exp\left\{i \left( k - \frac{ut}{\ell^2}+ \frac{v}{D} \right)y \right\}, 
  \label{eq:iVSHE02}
\end{eqnarray}
%%%
where $L_y$ is the system dimension in the $y$ direction, and $k= 2 \pi n/L_y$ with integer $n$. Note that the above solution does not keep the vortex lattice structure, but it contains $N_{\rm v}= L_x L_y/2 \pi \ell^2$ vortices in the system where $L_x$ is the system dimension in the $x$ direction~\cite{Ruggeri76,Kato93}. Guided by the time dependence of $g_p (\bmr,t)$, we impose 
%%%
\begin{equation}
  \phi_k(t)= \phi_k(0) \exp(\Lambda k \ell^2 t), 
\end{equation}
%%%
where the above time dependence is justified only for a {\it linearlized} TDGL equation. With this note in mind, we have 
%%%
\begin{subequations} \label{all}
  \begin{align} 
    \partial_t \phi_k &= \Lambda k \ell^2 \phi_k \label{eq:DtDphi01} \\ 
    \partial_t u_k &= \left[ \frac{u (x+k \ell^2)}{\ell^2}
      - i \frac{uy}{\ell^2}  \right] u_k, \label{eq:DtDu01} \\
    Q_x^2 u_k &= \left[  \frac{1}{\ell^2}
      - \frac{(x+k \ell^2 -ut)^2}{\ell^4} \right] u_k,  \label{eq:DQxDu01} \\
    Q_y^2 u_k &= \left[ \frac{(x+k \ell^2 -ut)^2}{\ell^4}
      + \frac{2v(x+ k \ell^2)}{D\ell^2}  \right] u_k, \label{eq:DQyDu01}
  \end{align}   
\end{subequations}
%%%
where, as in the previous section, only terms up to the linear order with respect to $u$ and $v$ are collected, except for terms appearing in the $(x+k \ell^2 -ut)^2$ combination. 

Now substituting Eqs.~(\ref{eq:DtDphi01})--(\ref{eq:DQyDu01}) to Eq.~(\ref{eq:iVSHE01}) and using Eq.~(\ref{eq:TDGL04}), we obtain
\begin{eqnarray}
  && \Bigg\{ \left( \frac{u}{\ell^2}+ \Lambda + \frac{2v}{\ell^2}\right) (x+ k \ell^2) \nonumber \\
  && \qquad 
  + i \left( \frac{E_{\rm iVSHE}}{H_{\rm c2} \ell^2}- \frac{u}{\ell^2} \right)y
  + \eps+ \frac{D}{\ell^2} \Bigg\}
  \phi_k(t) u_k (t) = 0, \nonumber \\
\end{eqnarray}
%%%
from which we have two conditions, 
\begin{equation}
  u = \frac{E_{\rm iVSHE}}{H_{\rm c2}} 
  \label{eq:u02}
\end{equation}
%%%
and
\begin{equation}
  u+ \Lambda \ell^2 + 2v = 0, 
  \label{eq:uv02}
\end{equation}
%%%
at the immediate vicinity of $H_{\rm c2}$ determined by $\eps+ D/\ell^2=0$.

The values of $u$, $v$, and $E_{\rm iVSHE}$ are determined from the open circuit condition. To this end, we first evaluate the charge current ${\bm J}^{\rm (c)}$ carried by the vortex motion. As in the previous section, we can show that ${\bm J}^{\rm (c)}$ satisfies
%%%
\begin{equation}
  {\bm J}^{\rm (c)} = {\bm \nabla} \times \left( - K^{\rm (c)}
  |\Psi|^2 {\bf \hat{z}} \right)
  - \frac{ 2v K^{\rm (c)} }{D} |\Psi|^2 {\bf \hat{y}}.
  \label{eq:jc03}
\end{equation}
%%%
Then, after the spatial average we obtain 
%%%
\begin{equation}
  \bra {\bm J}^{\rm (c)} \ket_{\rm s} = 
  - \frac{2v K^{\rm (c)}}{D} \bra |\Psi|^2 \ket_{\rm s} {\bf \hat{y}}, 
  \label{eq:jc04}
\end{equation}
%%%
where the spatial average of $|\psi|^2$ is expressed as 
%%%
\begin{equation}
  \bra |\Psi|^2 \ket_{\rm s} = \frac{1}{N_{\rm v}} \sum_k |\phi_k(t)|^2. 
  \label{eq:Psi2ave01}
\end{equation}
%%%
The total charge current is given by 
%%%
\begin{equation}
  \bra {\bf J}^{\rm (c)}_{\rm tot} \ket
  =
  \sigma_{\rm n} E_{\rm iVSHE} {\bf \hat{y}}
  - \frac{2v K^{\rm (c)}}{D} \bra |\Psi|^2 \ket_{\rm s} {\bf \hat{y}}, 
\end{equation}
where $\sigma_{\rm n}$ is the normal-state electrical conductivity. Since $  \bra {\bf J}^{\rm (c)}_{\rm tot} \ket \parallel {\bf \hat{y}}$, the open circuit condition reads 
%%%
\begin{equation}
  \bra {\rm J}^{\rm (c)}_{{\rm tot}, y} \ket = \sigma_{\rm n} E_{\rm iVSHE}
  - \frac{2v K^{\rm (c)}}{D} \bra |\Psi|^2 \ket_{\rm s} = 0. 
  \label{eq:opencirc01}
\end{equation}
%%%
Substituting Eqs.~(\ref{eq:u02}) and (\ref{eq:opencirc01}) into Eq.~(\ref{eq:uv02}), we obtain
%%%
\begin{subequations} \label{all}
  \begin{align} 
  E_{\rm iVSHE} &= - \Lambda \ell^2 H_{\rm c2} \, \frac{\eta}{1+\eta}, \label{eq:E_iVSHE01} \\
  u &= - \Lambda \ell^2 \; \frac{\eta}{1+\eta}, \\
  v &= - \frac{\Lambda \ell^2}{2} \; \frac{1}{1+\eta},
  \end{align}   
\end{subequations}   
%%%
where the dimensionless parameter $\eta$ was defined in Eq.~(\ref{eq:eta01}). 

Next, we calculate the spin current ${\bm J}^{\rm (s)}$ carried by the vortex motion. Starting from Eq.~(\ref{eq:js01}) we find that the spin current {\it before the spatial average} is not perfectly parallel to the $x$ axis. Namely, in contrast to the {\it direct} vortex SHE, it has the $y$ component, 
%%%
\begin{subequations}   
  \begin{align}   
    J_x^{\rm (s)} &= - K^{\rm (s)} \frac{2 \sqrt{\pi \ell^2}}{L_y}
  \sum_{k,k'} \phi^*_k \phi_{k'} u^*_k u_{k'} \nonumber \\
    & \times \Bigg[ \frac{2u(x+ k \ell^2)(x+ k' \ell^2)}{\ell^4} \nonumber \\
    & \qquad + \Lambda k \ell^2(x+k' \ell^2) + \Lambda k' \ell^2(x+k\ell^2)
    \Bigg], \\
  J_y^{\rm (s)} &=  -i K^{\rm (s)} \frac{2 \sqrt{\pi \ell^2}}{L_y}
  \sum_{k,k'} \phi^*_k \phi_{k'} u^*_k u_{k'} \nonumber \\
  & \qquad \times \Bigg[ \Lambda k (x+ k' \ell^2) - \Lambda k'(x+ k\ell^2) \Bigg].
  \end{align}       
\end{subequations}   
%%%
{\it After the spatial average}, however, the $y$ component vanishes and the spin current becomes 
%%%
\begin{equation}
  \bra {\bm J}^{\rm (s)} \ket_{\rm s} = - u \frac{K^{\rm (s)}}{\ell^2} \bra |\Psi|^2 \ket_{\rm s} \; {\bf \hat{x}},
  \label{eq:js03}
\end{equation}
%%%
which coincides with Eq.~(\ref{eq:js02}), where $\bra |\Psi|^2 \ket_{\rm s}$ is now given by Eq.~(\ref{eq:Psi2ave01}). 

To summarize this section, we have developed a TDGL theory of the {\it inverse} vortex SHE. In particular, we have obtained the explicit expression of the induced transverse electric field within the TDGL theory, Eq.~(\ref{eq:E_iVSHE01}), for the first time. This phenomenon, in which an input spin accumulation gradient $\mu_1^{\rm (s)} {\bf \hat{x}}$ [or a spin current ${\bm j}^{\rm (s)}= (\sigma_{\rm n}/|e|) \mu_1^{\rm (s)} {\bf \hat{x}}$ accompanied by the vortex motion] drives the transverse charge current ($\parallel {\bf \hat{x}}$), can be viewed as a new member of the {\it inverse} SHE. This is because the phenomenology matches the definition of the {\it inverse} SHE~\cite{Takahashi08,Sinova15}. In this {\it inverse} vortex SHE, the moving Abrikosov lattice has a tendency to be broken by the driving force of spin accumulation gradient. We hence expect that effects of the nonlinear term of the TDGL equation may be important, which is left for future studies.

%%%%%%%%%%%%%%%%%%%%%%%%%%%%%%%%%%%%
\section{Discussion and Conclusion \label{Sec:VI}} 
%%%%%%%%%%%%%%%%%%%%%%%%%%%%%%%%%%%%

%%%%%%%%%%%%%%%%%%%%%%%%%%%%%%%%%%%%% 
\begin{figure}[t] 
  \begin{center}
    \includegraphics[width=6.5cm]{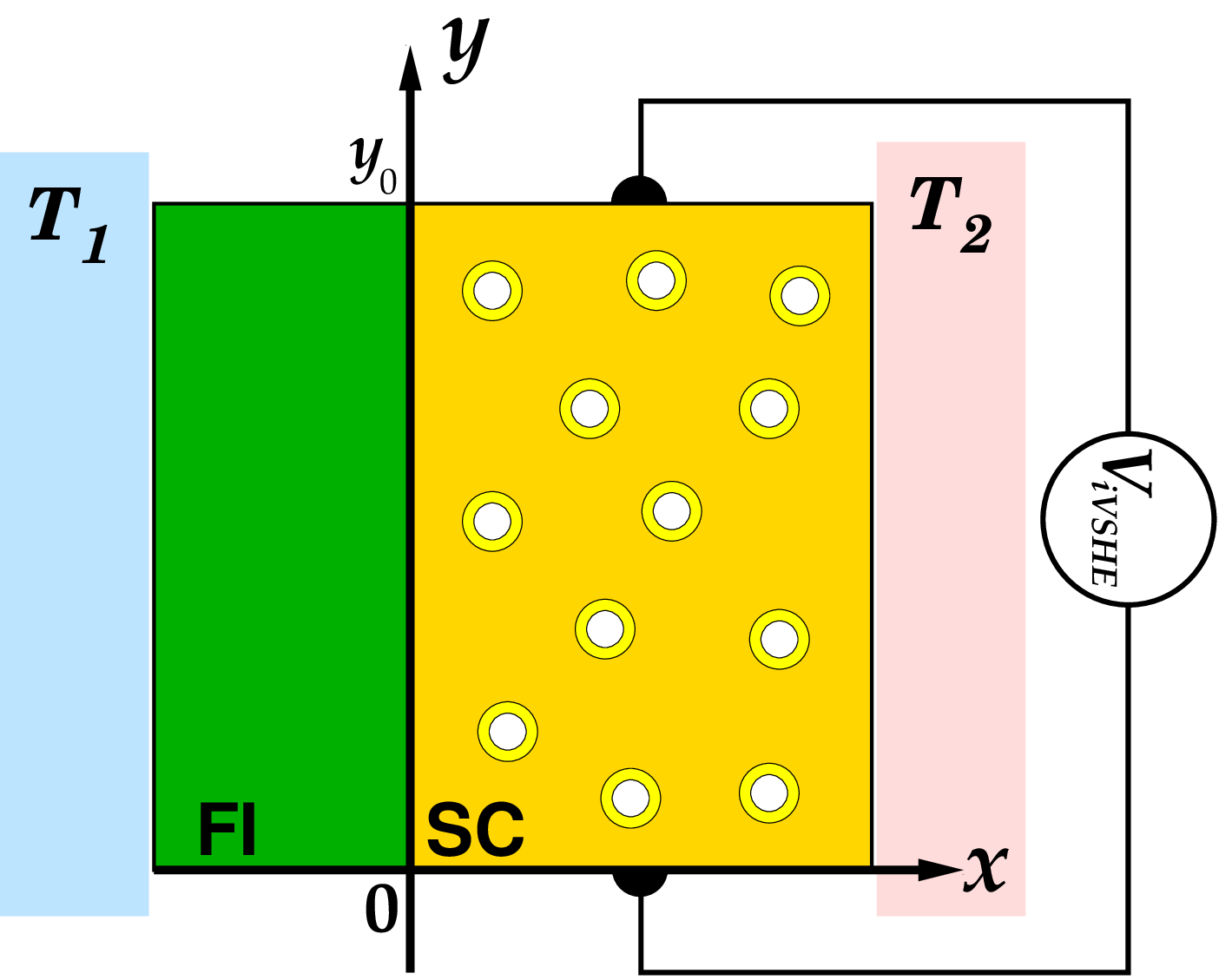}
  \end{center}
  \caption{Schematic illustration of the system discussed in Sec.~\ref{Sec:VI}. Here FI and SC refer to a ferromagnetic insulator and a type-II superconductor, respectively.} 
  \label{fig:VSHE-NE}
\end{figure} 
%%%%%%%%%%%%%%%%%%%%%%%%%%%%%%%%%%%%

In this paper, we have developed a TDGL theory of the {\it direct/inverse} vortex SHE. For the {\it direct} vortex SHE in which an input charge current drives the transverse spin current accompanied by the vortex motion, we have employed the well-known Schmid-Caroli-Maki solution~\cite{Schmid66,Caroli67} for the established TDGL equation under the applied electric voltage~\cite{Maki69}, and found out the expression of the induced spin current [Eq.~(\ref{eq:js02})]. For the {\it inverse} vortex spin Hall effect in which an input spin current drives the transverse charge current concomitant with the vortex motion, we have microscopically constructed the TDGL equation under the applied spin accumulation gradient [Eq.~(\ref{eq:TDGL02})], and calculated the induced transverse charge current [Eq.~(\ref{eq:jc04})] as well as the open circuit voltage [Eq.~(\ref{eq:E_iVSHE01})]. 
  
The impact of the present paper can be summarized as follows. First, we have explicitly written down the solution to the linearized TDGL equation for the {\it direct/inverse} vortex SHE. This provides us with an intuitive physical picture of the {\it direct/inverse} vortex SHE. Second, we have formulated the TDGL equation for the vortex SHE, which can be used for the numerical simulation of the {\it direct/inverse} vortex SHE. Indeed, as seen in literature, the TDGL equation has been widely used in the numerical simulation of the vortex states~\cite{Kato91,Al-Saidi03,Mukerjee04}, and therefore, the present paper can serve as a basis for a quantitative numerical simulation.

Before ending, we discuss how to detect the vortex SHE experimentally. To observe the vortex SHE, a convenient way is to focus on the {\it inverse} vortex SHE, and perform a spin injection via the spin Seebeck effect~\cite{Umeda18,Sharma23} or the spin pumping~\cite{Inoue17,Yao18,Rogdakis19,Jeon20}. It allows us to detect the {\it inverse} vortex SHE by measuring the open circuit voltage, and this procedure was actually adopted in Refs.~\cite{Umeda18,Sharma23}. Theoretically, giving a numerical estimate of the signal is an involved problem as the open circuit voltage [see Eq.~(\ref{eq:E_iVSHE01})] depends on the magnitude of the spin accumulation, which is quite hard to estimate. However one can discuss the {\it change} of the open circuit voltage from the normal to the vortex states (see Fig. 8 of Ref.~\cite{Taira21}). Experimentally, how the {\it inverse} vortex SHE manifests itself in the open circuit voltage is best seen in Ref.~\cite{Sharma23} (see Fig.~5 therein). In analyzing the data, care is necessary because there is a similarity between the {\it inverse} vortex SHE under discussion and the vortex Nernst effect~\cite{Wang06,Pourret06} as pointed out in the Introduction. Indeed, in a positive magnetic field region of Fig.~5b in Ref.~\cite{Sharma23}, a positive signal due to the vortex Nernst effect and a negative signal coming from the {\it inverse} vortex SHE, are clearly visible. This experiment, together with Ref.~\cite{Umeda18} proves the feasibility of the experimental detection of the vortex SHE. In passing, it is tempting to measure the {\it inverse} vortex SHE in the pseudo-gap phase of high-$T_{\rm c}$ cuprate superconductors~\cite{Wang06}, since the underlying physics of the {\it inverse} vortex SHE is similar to that of the vortex Nernst effect.

Below we discuss a way of identifying the {\it inverse} vortex SHE, by distinguishing it from the vortex Nernst effect from symmetry point of view. We consider the situation shown in Fig.~\ref{fig:VSHE-NE}, where a bilayer composed of ferromagnetic insulator (FI) layer and type-II superconductor (SC) is placed under a temperature bias $\Delta T= T_2 - T_1$. The situation is similar to that of Ref.~\cite{Umeda18,Sharma23} where the FI and SC correspond to Y$_3$Fe$_5$O$_{12}$ and NbN, respectively. Due to the spin Seebeck effect, spins are injected from FI to SC, and a spin accumulation gradient is established in the SC layer. Then, following the {\it inverse} vortex SHE discussed in Sec.~\ref{Sec:IVSHE}, a transverse voltage, 
%%%
\begin{equation}
  V_{\rm iVSHE} = \Lambda \ell^2 H_{\rm c2} \, \frac{\eta}{1+\eta} y_0 , 
\end{equation}
%%%
is generated, where $y_0$ is the sample dimension along the $y$ axis (Fig.~\ref{fig:VSHE-NE}). The crucial point in the above signal is that $V_{\rm iVSHE}$ contains the coefficient $\Lambda$, which is doubly proportional to $\mu^{\rm (s)}_0$ and $\mu^{\rm (s)}_1$ [see Eq.~(\ref{eq:Lambda01})]. The physics behind this is that for the {\it inverse} vortex SHE to occur, the vortex core needs to be spin polarized, and this requires the uniform component of the spin accumulation $\mu^{\rm (s)}_0$. Besides, the vortices need to be dragged by the spin accumulation gradient, which requires $\mu^{\rm (s)}_1$. Then, in the situation considered here, the spin accumulation $\mu^{\rm (s)}(\bmr)$ is caused by the spin Seebeck effect~\cite{Umeda18}, thereby both $\mu^{\rm (s)}_0$ and $\mu^{\rm (s)}_1$ are proportional to $\Delta T$. Consequently, $V_{\rm iVSHE}$ is proportional to the square of $\Delta T$, i.e.,  
%%%
\begin{equation}
  V_{\rm iVSHE} \propto (\Delta T)^2. 
  \label{eq:IVSHE_DT01}
\end{equation}
%%%
This signal is proportional to the square of the temperature bias, and thus the response in Eq.~(\ref{eq:IVSHE_DT01}) belongs to the nonlinear and nonreciprocal thermal transport~\cite{Tokura18,Nakai19}, and does not change sign under the operation $\Delta T \leftrightarrow - \Delta T$. By contrast, a parasitic voltage $V_{\rm VNE}$ coming from the vortex Nernst effect is proportional to $\Delta T$,
%%%
\begin{equation}
  V_{\rm VNE} \propto \Delta T.
  \label{eq:VNE_DT01}
\end{equation}
%%%
Using the symmetry difference between Eqs.~(\ref{eq:IVSHE_DT01}) and (\ref{eq:VNE_DT01}), we can disentangle the {\it inverse} vortex SHE from the vortex Nernst effect.

  The physics that a longitudinal vortex motion induces a transverse electric voltage is common to both the {\it inverse} vortex SHE and the vortex Nernst effect, where the difference is either the driving force is a spin accumulation gradient for the former, or a temperature gradient for the latter. Moreover, the same physics also applies to vortex motions driven by other means~\cite{Giaever65,Giaever66,Ekin74,He19,Xue19}. The new characteristic of the {\it inverse} vortex SHE is that the vortices carry spin polarization, which is different from previously known ones. Following Anderson~\cite{Anderson66}, all these transverse voltage generations can be understood using Fig.~1 of Ref.~\cite{Anderson66}, combined with the the Josephson equation [see Eq.~(13) therein]: 
%%%
  \begin{equation}
    \partial_t (\phi_1 - \phi_2) = 2|e| V_{12},
    \label{eq:josephson}
  \end{equation}
  %%%
  where $\phi_i$ $(i=1,2)$ is the phase of the pair field at point $i$ of Fig.~1 of \cite{Anderson66} and $V_{12}$ is the voltage drop between points $1$ and $2$. A vortex moves across a path from $1$ to $2$ is accompanied by a phase slippage [the left-hand side of Eq.~(\ref{eq:josephson})], thereby producing the transverse voltage [the right-hand side of Eq.~(\ref{eq:josephson})]. Note that when a vortex pinning~\cite{Reichhardt17} sets in, the phase of the pair field is frozen, i.e., the left-hand side of Eq.~(\ref{eq:josephson}) vanishes, and therefore the transverse voltage disappears. Based on this argument, an expected behavior of the open circuit voltage once taking account of the vortex pinning effect is drawn in Fig.~8(a) of Ref.~\cite{Taira21} (see the dotted line therein). Note also that, in the case of a vortex channel flow where only a bundle of vortices moves whereas the remaining vortices are pinned, the transverse voltage generation due to the inverse vortex SHE still exists. 

In summary, we have developed the TDGL theory of the vortex SHE. We have written down the explicit solutions of the TDGL equation for the {\it direct/inverse} vortex SHE, which provides us with an intuitive understanding of these phenomena. The TDGL theory of the {\it direct/inverse} vortex SHE presented here can be a basis for numerical simulations. Since this work have set up equations for numerical studies, we hope that the present study stimulates a more quantitative numerical approach in future. Moreover, the conventional SHE requires the spin-orbit interaction~\cite{Takahashi08,Sinova15}, but the vortex SHE investigated here does not rely on the spin-orbit interaction. Therefore, we also hope that future experiments on the vortex SHE using cuprate superconductors are performed as a realization of SHE {\it free from the spin-orbit interaction}.

%%%
\acknowledgments 
%%%
We are grateful to T. Taira for discussions in the early stage of this work. This work was financially supported by JSPS KAKENHI Grants (No. JP22H01941 and No. JP21H01799) and Asahi Glass Foundation. \\

\appendix

%%%%%%%%%%%%%%%%%%%%%%%%%%%%%%%%%%%%%%%%%%%%%%%
\section{TDGL equation with no driving force \label{Sec:App01}} 
%%%%%%%%%%%%%%%%%%%%%%%%%%%%%%%%%%%%%%%%%%%%%%

In the Appendix, we review how to construct the TDGL equation in the presence of the scalar potential gradient. There are several ways of microscopically deriving the TDGL equation~\cite{Caroli67,Ebisawa71}, but from the present perspective, the simplest way is perhaps to use the fact~\cite{Dominicis75} that the TDGL equation is derived from the Ginzburg-Landau action, 
%%%
\begin{eqnarray}
  S_{\rm GL} &=& \int_0^\beta d \tau \int d^3 r \Bigg\{ \Psi^*(\bmr, \tau) {\cal K}_2 (i \partial_\tau) \Psi (\bmr,\tau) \nonumber \\
  && + \frac{{\cal K}_4}{2} |\Psi(\bmr,\tau)|^4 \Bigg\}, 
  \label{eq:action01}
\end{eqnarray} 
%%%
as a stationary condition $\delta S_{\rm GL}/\delta \Psi^*=0$~\cite{Altland-Simons}, where $\tau$ is the imaginary time and $\Psi$ is the pair field, the latter of which is understood as a complex number when used within the functional integral representation of the partition function~\cite{Coleman-text} 
%%%
\begin{equation}
  {\cal Z} = \int {\cal D}[\Psi,\Psi^*] e^{-S_{\rm GL}}. 
\end{equation}
%%%
Writing down the above procedure more precisely, by firstly calculating ${\cal K}_2 (\omega_m)$ in the Matsubara space and then performing analytic continuation $\omega_m \to -i \omega = \partial_t$~\cite{Caroli67,Takayama70}, we obtain the TDGL equation from $S_{\rm GL}$ as 
%%%
\begin{equation}
  {\cal K}_2 (\partial_t) \Psi(\bmr,t)
  + {\cal K}_4 |\Psi(\bmr,t)|^2 \Psi(\bmr,t) = 0, 
  \label{eq:TDGL00}
\end{equation}
%%%
where $t= - i \tau$ is now the real time. For the moment we focus on the linearized TDGL equation, for which discussion of the Gaussian action is sufficient. Besides, we use the usual quasiclassical approximation~\cite{Werthamer-review} and employ the identity 
%%%
\begin{eqnarray}
  \exp{\Big( 2 i |e| \int_{\bmr'}^\bmr \bmA \cdot d \bms \Big)} \Psi(\bmr')
  = e^{ -i (\bmr- \bmr') \cdot \bmQ } \Psi(\bmr),
  \label{eq:gauge-grad01}
\end{eqnarray}
%%%
where $\bmQ = -i{\bm \nabla} + 2|e|\bmA$ is the gauge-invariant gradient. 

%%%%%%%%%%%%%%%%%%%%%%%%%%%%%%%%%%%%% 
\begin{figure}[t] 
  \begin{center}
    \includegraphics[width=7cm]{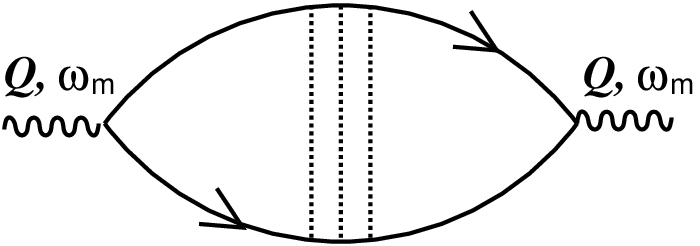}
  \end{center}
  \caption{Diagrammatic representation of the Gaussian action. A solid line with an arrow is electronic Green's function, a dotted ladder is the Cooperon, and a wavy line represents the pair field.} 
  \label{fig:diagram01}
\end{figure}
%%%%%%%%%%%%%%%%%%%%%%%%%%%%%%%%%%%%

Let us first see how the above procedure works in the derivation of the TDGL equation~\cite{Caroli67,Takayama70} in the absence of the driving force. In the dirty limit, the kernel ${\cal K}_2$ is calculated as 
%%%
\begin{eqnarray}
  {\cal K}_2 (\omega_m) &=& \frac{1}{|g|}- {\cal P} (\omega_m) ,
  \label{eq:K2_01}
\end{eqnarray}
%%%
where ${\cal P} (\omega_m)$ is the particle-particle polarization (see Fig.~\ref{fig:diagram01}) represented as 
%%%
\begin{eqnarray}
  {\cal P} (\omega_m) &=& \frac{T}{2} \sum_{\veps_n, \sigma} \int_\bmp
  G_{\bmp+ \bmQ, \sigma}(\veps_n+ \omega_m) G_{-\bmp, -\sigma}(-\veps_n) \nonumber \\
  && \times {\cal C}_{\bmQ, \sigma}(\veps_n+ \omega_m, -\veps_n),
  \label{eq:P01}
\end{eqnarray}
%%%
where $G_{\bmp, \sigma}(\veps_n)$ is the impurity-averaged Green's function defined in Eq.~(\ref{eq:Green01}), and ${\cal C}_{\bmQ, \sigma}$ is the Cooperon defined in Eq.~(\ref{eq:Cooperon01}). 

In the following, we consider terms up to the linear order with respect to the spatially-uniform component of the spin accumulation, $\mu_0^{\rm (s)}$. After the momentum integral, we obtain
%%%
\begin{eqnarray}
  {\cal P} (\omega_m) &=& 4 \pi N(0) T \sum_{\veps_n > 0}^{\omega_D} \frac{1}{|2 \veps_n+ \omega_m|+ DQ^2},
  \label{eq:P04}
\end{eqnarray}
%%%
where $\omega_D$ is the Debye frequency giving the cutoff. Note that, after the spin summation, the dependence of ${\cal P}$ on the spin accumulation becomes higher order than $\mu_0^{\rm (s)}$-linear term and hence discarded. In performing the resultant Matsubara frequency summation, we use the relation $4 \pi T \sum_{\veps_n > 0}^{\omega_D} (2 \veps_n + u)^{-1}= \ln (\omega_D/2 \pi T)- \varPsi(1/2+ u/4 \pi T)$, where $\varPsi(z)$ is the digamma function. Then, after performing the Matsubara frequency summation and analytic continuation $\omega_m \to -i \omega$, we have 
%%%
\begin{equation}
  {\cal K}_2 (-i \omega) =
  N(0) \Bigg[ \ln \left( \frac{T}{T_c}\right) + \frac{\pi}{8T_c} (-i \omega+ DQ^2) \Bigg],  
\end{equation}
%%%
where we used $1/(|g|N(0)) = \ln (\omega_D/2 \pi T_c)- \varPsi(1/2)$. In the above equation, we expanded the result to linear order in $-i \omega+ DQ^2$ and used $\varPsi^{(1)}(1/2)= \pi^2/2$, where $\varPsi^{(n)}(z)= d^n \varPsi(z)/d z^n$. Substituting the above result into Eq.~(\ref{eq:TDGL00}) and setting $-i \omega= \partial_t$~\cite{Caroli67,Takayama70}, we reproduce the well-known TDGL equation: 
%%%
\begin{equation}
  \left( \frac{\pi}{8T_c} \partial_t + \frac{T-T_c}{T_c} + \xi_0^2 \bmQ^2 + b |\Psi(\bmr,t)|^2 \right) \Psi(\bmr,t)= 0,
  \label{eq:TDGL01}
\end{equation}
%%%
where $\xi_0^2= \pi D/8 T_c$. Note that the coefficient of the nonlinear term coming from ${\cal K}_4$, $b= 7 \zeta(3)/8\pi^2T_c^2$, is added, which can be derived in a standard manner~\cite{Werthamer-review}.

%%%%%%%%%%%%%%%%%%%%%%%%%%%%%%%%%%%%% 
\begin{figure}[t] 
  \begin{center}
    \includegraphics[width=8.5cm]{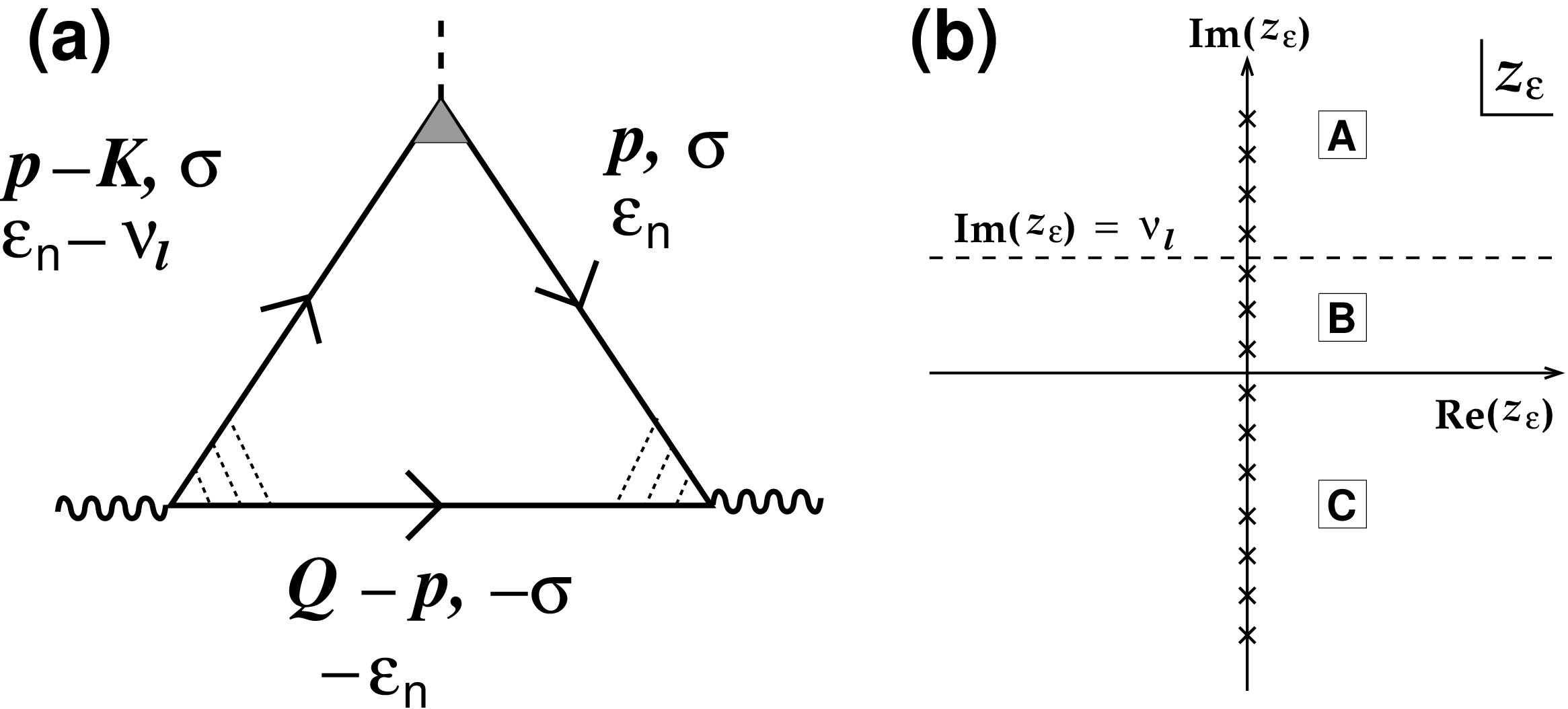}
  \end{center}
  \caption{(a) Diagrammatic representation of the contribution from the scalar potential gradient to the Gaussian action. The dashed line is the scalar potential gradient [Eq.~(\ref{eq:scalar01})], and the hatched triangle represents the diffuson [Eq.~(\ref{eq:diffuson01})]. (b) Three regions of Matsubara summation in the complex plane, leading to Eq.~(\ref{eq:dltP02}). }
  \label{fig:diagram03}
\end{figure} 
%%%%%%%%%%%%%%%%%%%%%%%%%%%%%%%%%%%%

%%%%%%%%%%%%%%%%%%%%%%%%%%%%%%%%%%%%%%%%%%%%%%%
\section{TDGL equation under scalar potential gradient \label{Sec:App02}} 
%%%%%%%%%%%%%%%%%%%%%%%%%%%%%%%%%%%%%%%%%%%%%%

The effect of the scalar potential gradient on the TDGL equation has been discussed in Ref.~\cite{Takayama70}. Here, we review their procedure of deriving the TDGL equation under the scalar potential gradient. We consider a response to the following external Hamiltonian: 
%%% 
\begin{eqnarray}
  {\cal H}^{\rm (c)}_{\rm ext} &=& -|e| \sum_{\sigma} \int d^3 r \, V^{\rm (c)} (\bmr, \tau) 
      \psi^\dag_\sigma (\bmr) \psi_\sigma (\bmr), 
  \label{eq:H_ext02}  
\end{eqnarray}
%%%
where the spatially uniform part of $V^{\rm (c)}$ is absorbed into the definition of the Fermi energy. In the above expression of the scalar potential, as discussed in Ref.~\cite{Takayama70}, we first need to keep the Matsubara frequency dependence as 
%%%
\begin{equation}
  V^{\rm (c)}(\bmr, \tau)= V^{\rm (c)}_\bmK e^{i (\bmK \cdot \bmr - \nu_l \tau)},
  \label{eq:scalar01}
\end{equation}
%%%
where $V^{\rm (c)}_\bmK e^{i \bmK \cdot \bmr} = -E x$ is the scalar potential, $V^{\rm (c)}_\bmK= iE \partial_{k_x}$, and the limit $\bmK \to 0$ is taken. We next perform the analytic continuation to real frequency $\nu_l \to -i \nu$, and set $\nu \to 0$ in the final step.

Then, in response to ${\cal H}^{\rm (c)}_{\rm ext}$, the kernel ${\cal K}_2$ [Eq.~(\ref{eq:K2_01})] is modified as
%%%
\begin{eqnarray}
  {\cal K}_2 (\omega_m) &=& \frac{1}{|g|}- \Bigg( {\cal P} (\omega_m) + \delta {\cal P}  \Bigg), 
    \label{eq:K2_02}
\end{eqnarray}
%%%
where $\delta {\cal P}$ is the first-order perturbation of the particle-particle polarization. Then the corresponding expression of $\delta {\cal P}$ is given by [see Fig.~\ref{fig:diagram03}(a)], 
%%%
\begin{eqnarray}
  \delta {\cal P} &=& T \sum_\sigma \sum_{\veps_n} \int_\bmp
  G_{\bmp, \sigma}(\veps_n ) G_{\bmp-\bmK, \sigma}(\veps_n) G_{\bmQ-\bmp, -\sigma}(-\veps_n) 
  \nonumber \\
  &\times&   V_\bmK^{\rm (c)} e^{i \bmK \cdot \bmr- i \nu_l \tau }
  {\cal C}_{\bmQ, \sigma}(\veps_n - \nu_l , -\veps_n)
  {\cal C}_{\bmQ, \sigma}(\veps_n , -\veps_n)  \nonumber \\
  &\times& {\cal D}_{\bmK} (\veps_n- \nu_l, \veps_n) ,
  \label{eq:dltP01}
\end{eqnarray}
%%%
where ${\cal D}_\bmK$ represents the diffuson, 
%%%
\begin{eqnarray}
  {\cal D}_{\bmK} (\veps_n- \nu_l, \veps_n)  \hspace{4cm} &&\nonumber \\
  = 
  \begin{cases}
    \frac{ \tau_{\rm imp}^{-1} }
         {|\nu_l|+D K^2}
         & \text{if $\veps_n(\veps_n - \nu_l)<0 $,} \\
         1                 & \text{otherwise.} 
  \end{cases} &&
  \label{eq:diffuson01}  
\end{eqnarray}
%%%
In performing the Matsubara frequency summation, there are three regions in the Matsubara space separated by two branch cuts as shown in Fig.~\ref{fig:diagram03}(b). In contrast to the case with the spin accumulation gradient discussed in Sec.~\ref{Sec:II}, two contributions from regions \fbox{A} and \fbox{C} cancel in the limit of $\nu \to 0$, and the dominant contribution comes from \fbox{B}.

Then, performing the momentum integral as well as the Matsubara frequency summation, we have
%%%
\begin{eqnarray}
  \delta {\cal P} &=&  N(0) \left( \frac{-i \nu}{-i \nu+ DK^2} \right)
  \frac{ i V^{\rm (c)}_\bmK e^{i \bmK \cdot \bmr}}{2 \pi T_c}
  \varPsi^{(1)} \left(\frac{1}{2} \right), \nonumber \\
  &=& N(0) \frac{i \pi V^{\rm (c)} (\bmr) }{4 T_c}, 
  \label{eq:dltP02} 
\end{eqnarray}
%%%
where, in moving to the second line, we used $\bmK^2 V^{\rm (c)}_\bmK=0$ ($ \because {\rm div} \bmE = 0$) and $\varPsi^{(1)}(1/2)= \pi^2/2$.

Then, substituting Eq.~(\ref{eq:dltP02}) into Eq.~(\ref{eq:K2_02}), setting $\omega_m \to -i \omega= \partial_t$~\cite{Caroli67,Takayama70}, and using Eq.~(\ref{eq:TDGL00}), we finally obtain the TDGL equation under the scalar potential gradient: 
%%%
\begin{eqnarray}
  && \Bigg( \frac{\pi}{8T_c} \big(\partial_t - 2 i |e|V^{\rm (c)}(\bmr) \big)+ \frac{T-T_c}{T_c} + \xi_0^2 \bmQ^2 \Bigg) \Psi(\bmr,t) \nonumber \\
  && \hspace{3cm} + b |\Psi(\bmr,t)|^2 \Psi(\bmr,t)= 0. 
  \label{eq:TDGL01c}
\end{eqnarray}
%%%


%\clearpage
%\subsection*{Figure Captions} 

\end{document}